\documentstyle[12pt,epsfig]{article}
\def\slash#1{\setbox0=\hbox{$#1$}#1\hskip-\wd0\hbox to\wd0{\hss\sl/\/\hss}}
\begin{document}
%\magnification=1200
\baselineskip=20 pt
\def\bea{\begin{eqnarray}}
\def\eea{\end{eqnarray}}
\def\g{\gamma}
\def\mphi{m_{\phi}}
\def\vphi{<\phi>}
\def\s{\sigma}
\def\m{\mu}
\def\n{\nu}
\def\mmu{m_{\mu}}
\def\L{\Lambda}
\def\ep{\epsilon}
\begin{center}
{\Large{\bf Testable $(g-2)_{\mu}$ contribution due to a
light

stabilized radion in the Randall-Sundrum model}}
\end{center}

\vskip 10pT
\begin{center}
{\large\sl \bf{Prasanta Das}~\footnote{E-mail: pdas@iitk.ac.in}
}
\vskip  5pT
{\rm
Department of Physics, Indian Institute of Technology, \\
Kanpur 208 016, India.} \\
\end{center}

%\vskip 0.02in
\begin{center}
{\large\sl  \bf{Uma Mahanta}~\footnote{E-mail:mahanta@mri.ernet.in}
}
\vskip 5pT
{\rm
Mehta Research Institute, \\
Chhatnag Road, Jhusi
Allahabad-211019, India .}\\
\end{center}  

\vspace*{0.02in}

\begin{center}
{\bf Abstract}
\end{center}
In this paper we calculate the $(g-2)_{\mu}$ contribution due to a light
stabilized radion using the radion couplings both to the kinetic
energy and the mass term of the muon. We find that the $(g-2)_{\mu}$
contribution due to radion diverges logarithmically with the cut off.
We then show that the bound from precision EW data on radion phenomenology
allows a sizable shift in the radion mediated muon anomaly that could be 
detected or tested with the present precision and certainly with the future
precision for measuring muon anomaly.

\newpage
Recently there has been a lot of interest in studying the phenomenology
of models of large \cite{ADD} and small \cite{RS} 
extra dimensions.
 Various phenomenological data have been 
used to put bounds on the unknown parameters of both large and small extra
dimensions. In particular the precision electroweak (EW) data \cite {PR}
and the (g-2) value of the muon \cite {MG} have been used to constrain
these parameters.  In this
paper we calculate the $(g-2)_{\mu}$ due to a stabilized radion \cite{GW}
in the Randall-Sundrum model. Using the radion couplings both to the
kinetic energy and mass term of the muon we find that the radion
contribution to the muon
anomaly diverges logarithmically with the cut off. For a light radion with
a mass of few tens of GeV and a radion vev of around a TeV, we obtain a muon
anomaly of the order of a few times $10^{-9}$. The values of $\mphi$
and $\vphi$ used by us in arriving at our numerical results are
chosen so as to be consistent
with the bounds implied by precision EW data on radion phenomenology.

The Feynman diagrams that give rise to the radion contribution to the muon
anomaly are shown in Fig 1.

The Feynman rules that are necessary for evaluating these diagrams can
be found in Ref \cite{DM1}.  We find that

\bea
I_{a}=-{e\over \vphi^2}\int {d^4l\over (2\pi )^4}{[{3\over 2}(\slash l
+ \slash p +2 \slash q )
-4\mmu ](\slash l + \slash q) +\mmu )\slash {\ep} (\slash l +\mmu )
[{3\over 2}(\slash l +\slash p) -4\mmu ]
\over [(l+q)^2-\mmu^2][l^2-\mmu^2][(l-p)^2-\mphi^2]}
\eea

and

\bea
I_{b}={3e\over \vphi^2}\int {d^4l\over (2\pi )^4} {[{3\over 2}
 (\slash l +\slash p +\slash q) -4\mmu ](\slash l +\mmu )\slash {\ep}\over 
(l^2-\mmu ^2)[(l-p-q)^2-\mphi^2]}
\eea

The  expressions for the loop integrals $I_a$ and $I_b$ given above
arise from Figs 1a and 1b respectively. 

\vspace*{-2.5in}
%%%%%%%%%%%%%%%%%%%%%%%%%%%%%%%%%%%%%%%%%%%%%%
\newpage
\begin{figure}
\begin{center}
\epsfig{file=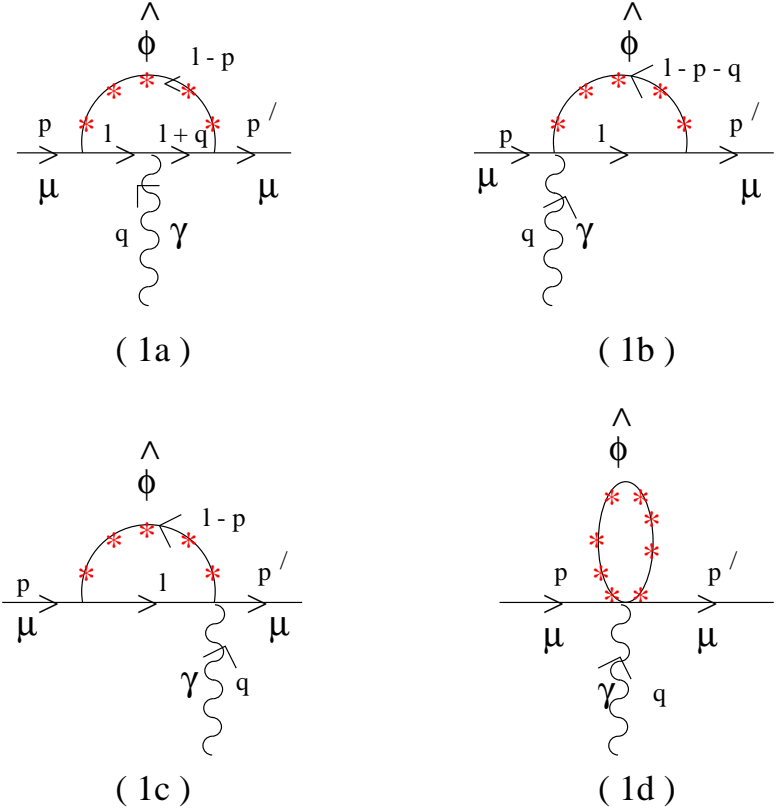}
\end{center}
\caption{\it { Feynman diagrams that gives rise to the radion contribution
to muon anomaly.}}
\end{figure}
%%%%%%%%%%%%%%%%%%%%%%%%%%%%%%%%%%%%%%%%%%%%%%%
$\ep^{\mu}(q)$ is the photon polarization vector. We have chosen the
incoming muon momentum p and the photon momentum q to express $I_{a}$ and
$I_{b}$. It can be shown that Fig 1c
and Fig 1d do not give rise to a term proportional to $q^{\nu}\s_{\m\n}$.
These two diagrams therefore
 do not contribute to the anomalous magnetic moment
of the muon and can be omitted from further discussion.
The contributions of Figs 1a and 1b to the anomalous magnetic moment of the
muon can be shown to be given by

\bea
I_{a}=-{9ie\over 2\vphi^2}\mmu \ep ^{\mu}(q) q^{\n}\s_{\m\n}\int x dx dy\int
{d^4l\over (2\pi )^4} {l^2\over D_a}+....
\eea

and 
\bea
I_b= {3ie\over 2\vphi^2}\mmu \ep ^{\mu}(q)
 q^{\n}\s_{\m\n}\int (3-5x)dx {1\over D_b}
+...
\eea

where

\bea
D_a= l^2-(1-x)^2q^2+2xy(1-x)p.q+-x^2y^2p^2+(1-x)^2q^2
\nonumber\\
+p^2xy
-\mmu^2(1-xy)-\mphi^2xy
\eea

and 
\bea
D_b=(l-x(p+q))^2+x(1-x)(p+q)^2-x\mphi^2-(1-x)\mmu^2
\eea

In the above we have dropped terms proportional to $\mmu^3 q^{\n}\s_{\m\n}$
from $I_a$ and $I_b$. The contributions arising from these terms will 
be suppressed compared to those that are proportional $\mmu$ for 
$\mphi^2\gg \mmu^2$. Further
 in this paper we shall be interested only in the static
values of the muon magnetic moment. 
In this static  or low energy approximation
we can set $p^2=p^{\prime 2}=p.q=0$ in the denominator after the magnetic
moment term proportional to $q^{\n}\s_{\m\n}$ has been extracted out from the
numerator.

In this static approximation we get

\bea
I_a \approx {9e\mmu \epsilon^\mu(q) \over 64\pi^2\vphi^2}q^{\n}\s_{\m\n}
\ln {\L^2\over\mphi^2}+...
\eea

and 

\bea
I_b \approx {-3e\mmu \epsilon^\mu(q) \over
64\pi^2\vphi^2}q^{\n}\s_{\m\n}(\ln 
{\L^2\over\mphi^2}+ {5\over 2})+...
\eea
where $\L$ is an ultraviolet momentum cut off. In the Randall-Sundrum model
the cut off $\L$ 
 can be identified with the mass of the lightest Kaluza-Klein
mode of the graviton in the several TeV range.
 To arrive at the above result we have assumed that $\mphi^2\gg\mmu^2$.
The above contributions to the muon magnetic moment can be put in the form
of an effective Lagrangian
\bea
L_{eff}\approx {3e\mmu\over 32\pi^2 \vphi^2}\partial^{\n}A^{\m}
\bar{\psi}\s_{\m\n}\psi (\ln {\L^2\over \mphi^2}-{5\over 4})
\eea

The radion contribution to the muon magnetic moment is therefore given by
\bea
a^r_{\m}\approx {3\mmu^2\over 16\pi^2\vphi^2}(\ln {\L^2\over \mphi^2}-
{5\over 4})
\eea

The UV cut off $\L$ for low energy radion phenomenology can be estimated
by using naive dimension analysis (NDA) \cite{GM}
The NDA estimate stipulates
that $\L =4\pi \vphi$, since ${1\over\vphi}$ acts as the expansion
parameter for non-renormalizable radion couplings to muon. In general
however the 
cut off $\L$ can be related to $\vphi$ via $\L =k \vphi$ where k lies between
1 and 4$\pi$. In the numerical results presented in this paper we shall 
take k to be equal to the geometric mean of 1 and 4$\pi $. We shall
also ensure that the values of $\mphi$ and $\vphi$ used to estimate
$a^r_{\m}$ satisfies the precision EW constraints.

\newpage 
The oblique EW parameter T has been used to put bounds on $\mphi$ and
$\vphi$ \cite{DM2}. 
These bounds can be represented in terms of an allowed region
and forbidden region in the $\mphi -\vphi$ plane [see Fig 2].
%%%%%%%%%%%%%%%%%%%%%%%%%%%%%%%%%%%%%%%%%%%%%
\begin{figure}[h]
\begin{center}
\epsfig{file=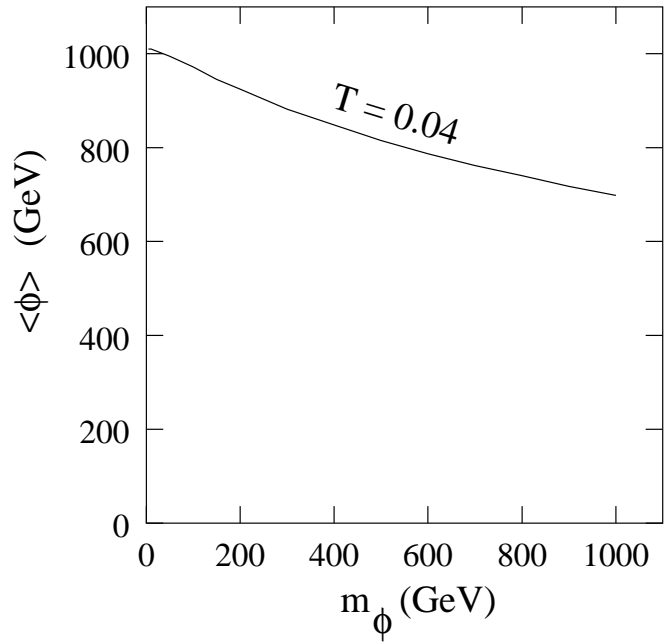}
\end{center}
\caption{\it {$\rho$ parameter constraints on radion vev $\langle \phi
\rangle$ and radion mass $m_\phi$. The allowed region lies above the
curve}}
\end{figure}
%%%%%%%%%%%%%%%%%%%%%%%%%%%%%%%%%%%%%%%%%%%%%%

For a light radion with a mass of 10 Gev, the precision EW constraint
forces the radion mediated muon anomaly to be less than or equal to 
2.7 $\times 10^{-9}$. On the other hand for a heavy radion with
a  mass of 500 Gev, the T parameter constraint on radion phenomenology
allows a muon anomaly of 1.5 $\times 10^{-9}$ as shown in Fig. 3.
%%%%%%%%%%%%%%%%%%%%%%%%%%%%%%%%%%%%%%%%%%%
\newpage
\begin{figure}
\begin{center}
\epsfig{file=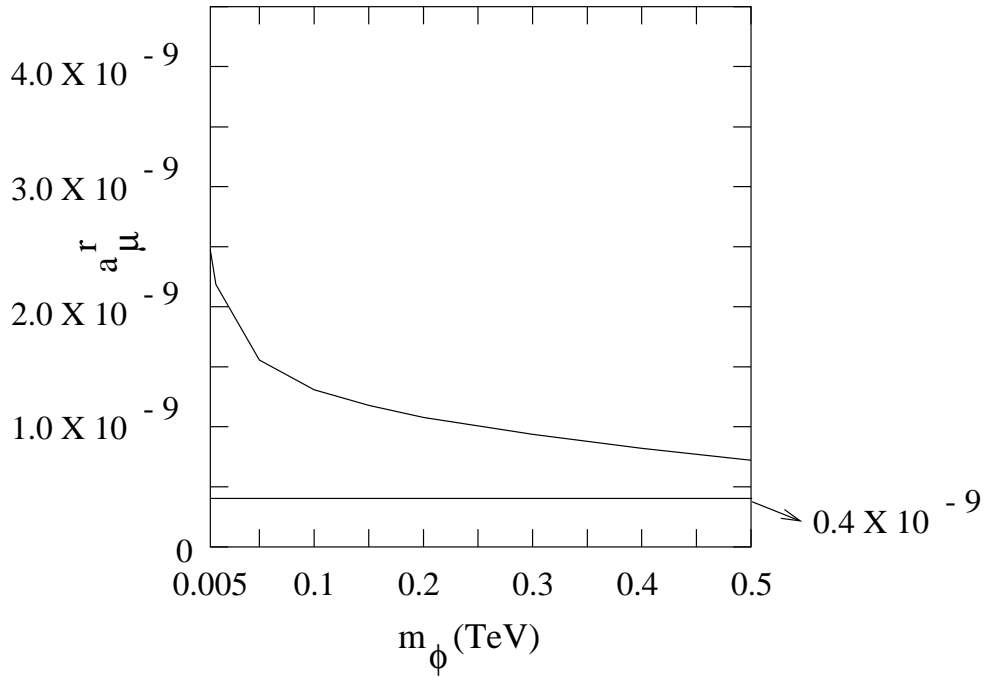}
\end{center}
\caption{\it {Plot of muon anomaly $a_\mu^r$ against the radion mass
$m_\phi$ (TeV).The horizontal line corresponds to the ultimate
precision
of the experiment.}}
\end{figure}
%%%%%%%%%%%%%%%%%%%%%%%%%%%%%%%%%%%%%%%%%%%%%%%%%%%%%%%
We would like to note that the log divergence of our result arises from
the radion coupling  to the kinetic energy term of the muon
which gives rise to a stronger divergence to the loop integral.
 Previous
estimates of radion mediated muon anomaly used the radion coupling
only to the mass term of the muon. In fact their radion coupling to
fermion is similar to that of the Higgs boson.
 Therefore they do not get
the log divergence. Actually they get a subdominant contribution
proportional to $\mmu^4$ which is correct for the Higgs boson but
not for the radion. It can be shown that the
 radion couplings to the muon reduces to the mass 
term of the muon only if both the muon lines are on shell. However in 
calculating the loop diagrams
 shown in Fig 1 one certainly cannot assume that
 muon lines at each vertex
 are on shell and hence their result is not trustable.

The Muon $(g-2)_{\mu}$ collaboration has reported a new improved measurement
of positive muon anomaly \cite{HNB}

$$a_{\mu}(expt)=(11659202 \pm 14 \pm 6)\times 10^{-10}$$

The muon anomaly expected in the SM according to the latest calculations
is given by

$$a_{\mu}(SM)=(11659176.96 \pm 6.4)\times 10^{-10}$$

This shows a discrepancy from the experimental value given by
$\delta a_{\mu}=(26 \pm 16 )\times 10^{-10}$. The ultimate goal of the 
Collaboration is to reduce the error to 4 $\times 10^{-10}$. In this paper
we have shown that the T parameter constraint on $\mphi$ and $\vphi$
gives rise to a radion mediated muon anomaly which is of the same
order as the present precision ($1.5\times 10^{-9}$)
 for measuring the muon anoamly. However
 with the ultimate precision 
of the experiment the level of muon anomaly presented in this paper
can certainly be detected or tested.

\end{document}